\documentclass{aastex631}

\usepackage{color}
\usepackage{amsmath}
\usepackage{changepage}
\usepackage{float}

\begin{document}

\title{Observations of Locally Excited Waves in the Low Solar Atmosphere Using the Daniel K. Inouye Solar Telescope (DKIST)}

\correspondingauthor{Shah Mohammad Bahauddin}
\email{shahmohammad.bahauddin@colorado.edu}

\author{Shah Mohammad Bahauddin}
\affiliation{Laboratory for Atmospheric and Space Physics, University of Colorado, Boulder, CO 80303, USA}
\affiliation{DKIST Ambassador, National Solar Observatory, Boulder, CO 80303, USA}

\author{Catherine E. Fischer}
\affiliation{National Solar Observatory, Boulder, CO 80309, USA}

\author{Mark P. Rast}
\affiliation{Department of Astrophysical and Planetary Sciences, University of Colorado, Boulder, CO 80309, USA}
\affiliation{Laboratory for Atmospheric and Space Physics, University of Colorado, Boulder, CO 80303, USA}

\author{Ivan Milic}
\affiliation{Leibniz-Institut für Sonnenphysik, Freiburg im Breisgau, Germany}

\author{Friedrich Woeger}
\affiliation{National Solar Observatory, Boulder, CO 80309, USA}

\author{Matthias Rempel}
\affiliation{High Altitude Observatory, NSF National Center for Atmospheric Research, Boulder, CO 80307, USA}

\author{Peter H. Keys}
\affiliation{Astrophysics Research Centre, School of Mathematics and Physics, Queen’s University Belfast, Belfast, BT7 1NN, UK}

\author{Thomas R. Rimmele}
\affiliation{National Solar Observatory, Boulder, CO 80309, USA}



\begin{abstract}


We present an interpretation of the recent Daniel K. Inouye Solar Telescope (DKIST) observations of propagating wavefronts in the lower solar atmosphere. Using MPS/University of Chicago MHD (MURaM) radiative magnetohydrodynamic simulations spanning solar photosphere, overshoot region, and lower chromosphere, we identify three acoustic-wave source mechanisms, each occurring at a different atmospheric height.  We synthesize the DKIST Visible Broadband Imager (VBI) G-band, blue-continuum, and Ca\,II\,K signatures of these waves at high spatial and temporal resolution, and conclude that the wavefronts observed by DKIST likely originate from acoustic sources at the top of the solar photosphere overshoot region and in the chromosphere proper. The overall importance of these local sources to the atmospheric energy and momentum budget of the solar atmosphere is unknown, but one of the excitation mechanism identified (upward propagating shock interaction with down-welling chromospheric plasma resulting in acoustic radiation) appears to be an important shock dissipation mechanism. Additionally, the observed wavefronts may prove useful for ultra-local helioseismological inversions and promise to play an important diagnostic role at multiple atmospheric heights.




\end{abstract}



\section{Introduction} \label{sec:intro}

Investigating the low solar atmosphere, from the deep photosphere to the low chromosphere, is central to understanding the transfer of mass and energy from the solar interior to the corona. Given the prevalence of acoustic waves in this region, understanding their source, propagation, and dissipation is crucial. Moreover, high resolution observations of these waves has significant diagnostic potential. 

The study of solar acoustic waves has a history of several decades, with theoretical analyses suggesting that these waves originate from discrete dynamical events both within and below the solar photosphere. They are responsible for the resonant global p-modes, and their high-frequency components propagate upward into the solar atmosphere, potentially contributing to momentum deposition and heating ~\citep[e.g.,][and references therein]{1946NW.....33..118B, 1948ApJ...107....1S, 2006ApJ...647L..73H, 2007ApJ...671.2154K, 2020A&A...642A..52A, 2021A&A...652A..43Y, 2023ApJ...945..154M}. Various mechanisms for exciting solar acoustic waves have been proposed~\citep[e.g.,][]{1952RSPSA.211..564L, 1954RSPSA.222....1L, 1967SoPh....2..385S, 1991LNP...388..195S, 1994ApJ...423..474M, 1994ApJ...424..466G, 1995ApJ...443..863R, 1999ApJ...524..462R, 2001A&A...370..136S, 2007A&A...463..297S, 2019ApJ...872...34K, 2019ApJ...880...13Z, 2020MNRAS.495.4904Z, 2021A&A...656A..95P, 2022A&A...664A.164P}, with sporadic observational support for different mechanisms~\citep[e.g.,][]{1995ApJ...444L.119R, 1998MNRAS.298L...7C, 1998ApJ...495L..27G, 1999ApJ...516..939S, 2000ApJ...535..464S, 2000ApJ...535.1000S, 2001ApJ...561..444S, 2010ApJ...723L.134B, 2010ApJ...723L.175R, 2013JPhCS.440a2044L, 2020ApJ...901L...9L}. However, it remains unclear which mechanism is dominant, how well each couples to global p-modes, and which is most important for generating the high-frequency waves that contribute to atmospheric heating.

Accurate identification and characterization of solar acoustic sources necessitates careful separation of the source and the local wave field from the background convective flows and p-modes. This is difficult because the amplitudes of the individual sources and the waves they emit are typically much smaller than those of the granulation and the p-mode coherence patches. The signal-to-noise ratio (SNR) of the locally generated wavefield to the background is typically well below unity, and the spectral content of the source signal overlaps with that of the p-modes and, to some extent, the background granular motions. Furthermore, the spatial extent of the waves and their propagation speed dictate the need for imaging instruments with exceptionally-high spatial resolution ($< 40$ km) and high temporal cadence ($\Delta t < 5$ s). Consequently, direct observation of the local acoustic-source wavefield remains challenging and has only recently been achieved, in simulations by leveraging the time-scale difference between the source-induced propagating wavefronts and background contributions~\citep{2021ApJ...915...36B, 2023ApJ...955...31B} and in observations~\citep{2023SPD....5440703F} by employing the high-cadence high-resolution capabilities of the NSF's Daniel K. Inouye Solar Telescope (DKIST, \cite{2020SoPh..295..172R}).

During the DKIST Operations Commissioning Phase (OCP) Cycle 1, substantial effort was invested in taking high-resolution images of the low solar atmosphere with the Visible Broadband Imager (VBI, \cite{2021SoPh..296..145W}) in several wavelength channels. The nominal formation heights of the filtergrams acquired span from the base of the photosphere to the mid chromosphere. Some of the timeseries show evidence of propagating wavefronts~\citep{2023SPD....5440703F}, potentially originating from local acoustic sources. The wavefronts are not seen in a mid-photospheric channel (G-band) but are visible in a channel with contributions from the upper photosphere and lower chromosphere (Ca II K band).  As referenced above, most models of chromospheric waves have focused on waves generated below the photosphere, the high frequency tail of the solar p-modes, or because of their signature in the p-mode spectra\citep[e.g.,][]{1995A&A...299..245G, 1998ApJ...496..527R, 1998ApJ...495L.115N, 2000ApJ...535..464S}, on sources in the low photosphere (with the SDO/HMI Fe~I 617.3 nm and the SOHO/MDI Ni I 676.8 nm lines nominally forming at heights of 100 and 125 km above $\tau_{500}=1$ respectively~\citep{2011SoPh..271...27F}). In contrast, these new observations and the analysis presented in this paper reveal that numerous acoustic excitation events may occur much higher in the atmosphere, at the top of the photospheric overshoot region and in the low chromosphere.

We report here on propagating wavefronts observed in the DKIST/VBI Ca II K image timeseries that appear to emanate from local acoustic events. We support the interpretation of their origin as local by comparing the observed properties of the wavefronts with those of similar events found in MPS/University of Chicago Radiative MHD (MURaM) simulations of the lower solar atmosphere. Analogous wave sources in the simulated atmosphere are located in the upper photosphere and lower chromosphere, and forward modeled spectra of the DKIST/VBI observational channels confirm the visibility of the wavefronts generated by these sources in Ca II K filtergrams but not in the G-band. The combined evidence from observations and numerical simulations, suggests that acoustic-wave sources are ubiquitous in the low solar atmosphere.

\section{DKIST Observations} \label{sec:dkist}

\begin{figure*}[t!]
\centerline{\includegraphics[width=0.9\textwidth]{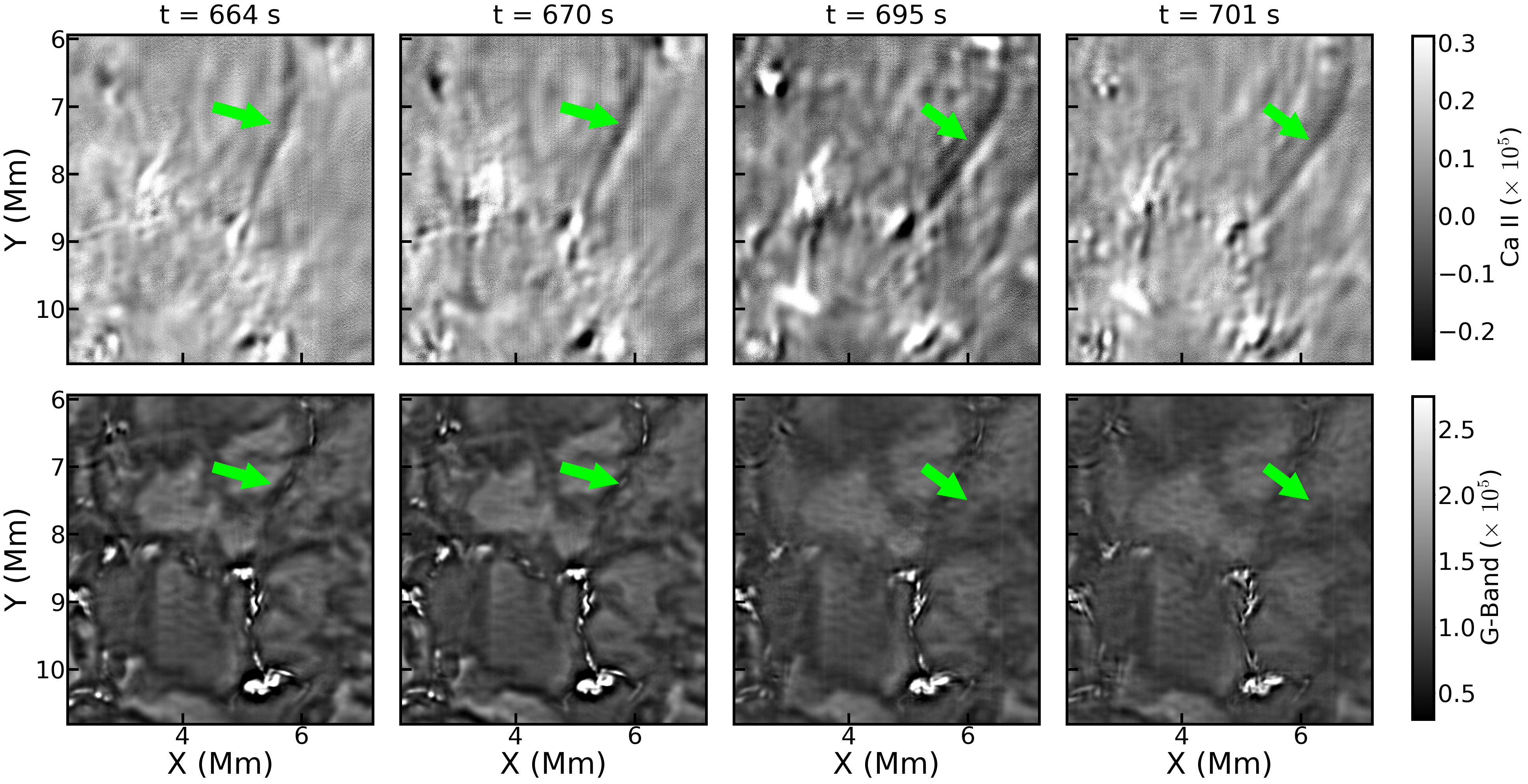}}
\caption{DKIST/VBI observations of wavefront propagation in the \ion{Ca}{2} K and corresponding G-band channels. The top row displays the time sequence of \ion{Ca}{2} K images with unsharp masking applied for contrast enhancement. The green arrow highlights the position of the wavefront. The bottom row presents the same time sequence in the G-band channel. No discernible wavefront is visible in G-band images even with unsharp masking, with the green arrows indicating where the wavefronts would be in the timeseries shown. This time series is delayed by 140 seconds relative to that of \ion{Ca}{2} K in order to account for the travel time of the waves between the photosphere and the low chromosphere.  Other delays were also examined (see text and Supplementary Video 1).} 
\label{fig:figure1}
\end{figure*}

We present an analysis of the quiet-sun image timeseries from the DKIST/VBI spanning 17:46 UTC to 19:31 UTC of 26 May 2022 (Dataset ID: AQKXV, AOLZW). The centers of the two regions observed (in helioprojective coordinates) lie at disk center, both with a field of view of $45.06\arcsec \times 45.06\arcsec$. The primary observations we employ are the timeseries taken in the \ion{Ca}{2} K (Center wavelength (CWL): $393.327$ nm, Full width at half maximum (FWHM): $0.101$ nm) and a G-band (CWL: $430.52$ nm, FWHM: $0.437$ nm) channels.

For these timeseries, the VBI camera collected bursts of images at 30 Hz, alternating between the \ion{Ca}{2} K and the G-band channels.  Short exposures, adaptive optics, and speckle reconstruction (based on the KISIP code of \cite{2008A&A...488..375W}) were combined to yield image time series with 6 s cadence, for \ion{Ca}{2} K and G-band respectively, and spatial sampling (pixel scale) of $0.011\arcsec$. The resulting images exhibit approximately $40$-pixel spatial shift between these two channels; this shift is corrected and a $17.6\arcsec \times 15.4\arcsec$ area is extracted for image post-processing and analysis. We note that these broad-band images are insensitive to the line-of-sight (LOS) velocity, so the wavefront motions we describe below are visible due to induced thermodynamic perturbations in the atmospheric regions contributing to each passband.

The strong \ion{Ca}{2} K resonant line we observe is frequently used for broadband characterization of solar and stellar activity. The line covers a wide range of wavelengths, with formation heights ranging from the photosphere to the mid-chromosphere (from $\tau_{500}=1$ to about 1500 km height). The \ion{Ca}{2} K channel of the DKIST/VBI is centered at the line ($393.327$ nm) with a width of 0.1 nm, and thus primarily samples the chromosphere, though it also includes contributions from the upper photosphere (down to a height of about 400 km, e.g., \cite{2010A&A...523A..55E}).

Opacity in the G-band arise from the absorption due to the presence molecules, primarily the CH (methylidyne) molecule. It exhibits larger contrast than pure continuum because the molecule number densities depend strongly on the temperature. With a passband of 0.437 nm, the VBI G-band channel probes photospheric regions from slightly above the blue continuum (a height of approximately 100\,km) to the temperature minima (e.g., \cite{2009LNP...778..129H}). G-band images typically capture fine details of the photosphere, including granular substructure and magnetic bright points. 

\subsection{Locally Excited Wavefronts in DKIST/VBI Data}

Inspection of the image timeseries captured by the DKIST VBI in the \ion{Ca}{2} K channel reveals spatially compact and temporally discrete propagating wavefronts~\citep{2023SPD....5440703F}. These transient events can be identified against the high amplitude, relatively slowly varying background flows because they are spatially resolved and because the temporal cadence of the observations is high enough to capture their propagation. The amplitudes of the transients are just marginally higher than the variance of the background granular flows and p-mode coherence patches (SNR is slightly over one). Consequently, the wavefronts are readily discernible when viewing image time-series but their identification and characterization in any single image is difficult. Because of the inherent time scale differences between the signal (the wavefront perturbation) and the background, temporal difference filtering can significantly improve wavefront SNR in simulations~\citep[][]  {2021ApJ...915...36B, 2023ApJ...955...31B}, but this method does not work well on this ground-based observational timeseries due to variations in image brightness between consecutive frames. Instead, we employ unsharp masking (IDL code \texttt{UNSHARP\_MASK}, for \ion{Ca}{2} K, \texttt{AMOUNT=20,  RADIUS=5} and for G-band, \texttt{AMOUNT=3,  RADIUS=5}) to enhance the contrast along the edges of wavefront perturbations. Such a contrast enhanced image timeseries is shown in Figure~\ref{fig:figure1}, with a \ion{Ca}{2} K images along the top row and G-band images along the bottom.  A video of the timeseries can be found in the supplementary materials online.

The green arrow in the upper row of images in Figure~\ref{fig:figure1} indicates the position of an observed wavefront propagating away from a source site in the \ion{Ca}{2} K time series. The wavefront has a width of about $0.4\arcsec$ ($\sim 300$ km), extends about 3 Mm, and propagates with a speed of about $14$ km/s.  Surprisingly, no discernible wave front is visible in co-aligned G-band images (bottom row in Figure~\ref{fig:figure1}), either co-temporally or staggered earlier or later in time. The G-band images were examined for the presences of propagating wavefronts for times spanning several minutes before and after their initial appearance in the \ion{Ca}{2} K images. This time frame (approximately $\sim$ $140$ seconds) is the approximate travel time of acoustic waves between the formation heights of the passbands, estimated by integrating the sound speed from the photospheric surface to a height of 1000 km above it in a realistically simulated photosphere (simulations details in Section \ref{sec:sim}). The investigation covers both forward and backward temporal directions due to the uncertainty regarding whether the wave initiates in the photosphere and then propagates to the chromosphere or vice versa. The absence of the wavefront in the G-band images is true of all \ion{Ca}{2} K events observed (an additional event is illustrated in the Supplementary Video 1), and suggests that the wavefront perturbations seen in the DKIST VBI \ion{Ca}{2} K data originate in the chromosphere and do not travel downward into or upward from the photosphere. This is supported by analysis of similar wavefronts found in magnetohydrodynamic (MHD) simulations.

\section{Locally Excited Wavefronts in a Simulated Atmosphere} \label{sec:sim}

We study a MURaM simulation of solar convection with physical dimension $L_x \times L_y \times L_z = 6.144 \times 6.144 \times 4.096$~Mm$^3$, where $L_z$ specifies the vertical direction. The domain has uniform $16$ km grid spacing in all directions. It extends for 1 hour of solar time, with output saved every $2.0625$ s. The data cube thus has the native dimensions $1800 \times 384 \times 384 \times 256$. The top boundary of the simulation is located 1.7 Mm above mean $\tau_{500} = 1$ and the depth of the convective portion of the layer below that is 2.3 Mm.  Horizontally periodic boundary conditions were employed in the solution, along with a semi-transparent upper boundary (closed for downflows and opened for upflows) and an open lower boundary (mass preserving). For reference, the simulation is a re-run of Simulation O16b from \cite{2014ApJ...789..132R}, with non-grey radiative transfer (four bins) and a domain extended an additional 1.024 Mm upwards.  The simulation computes the radiation field under the assumption of local thermodynamic equilibrium (LTE).

\begin{figure}[t!]
\centerline{\includegraphics[width=\linewidth]{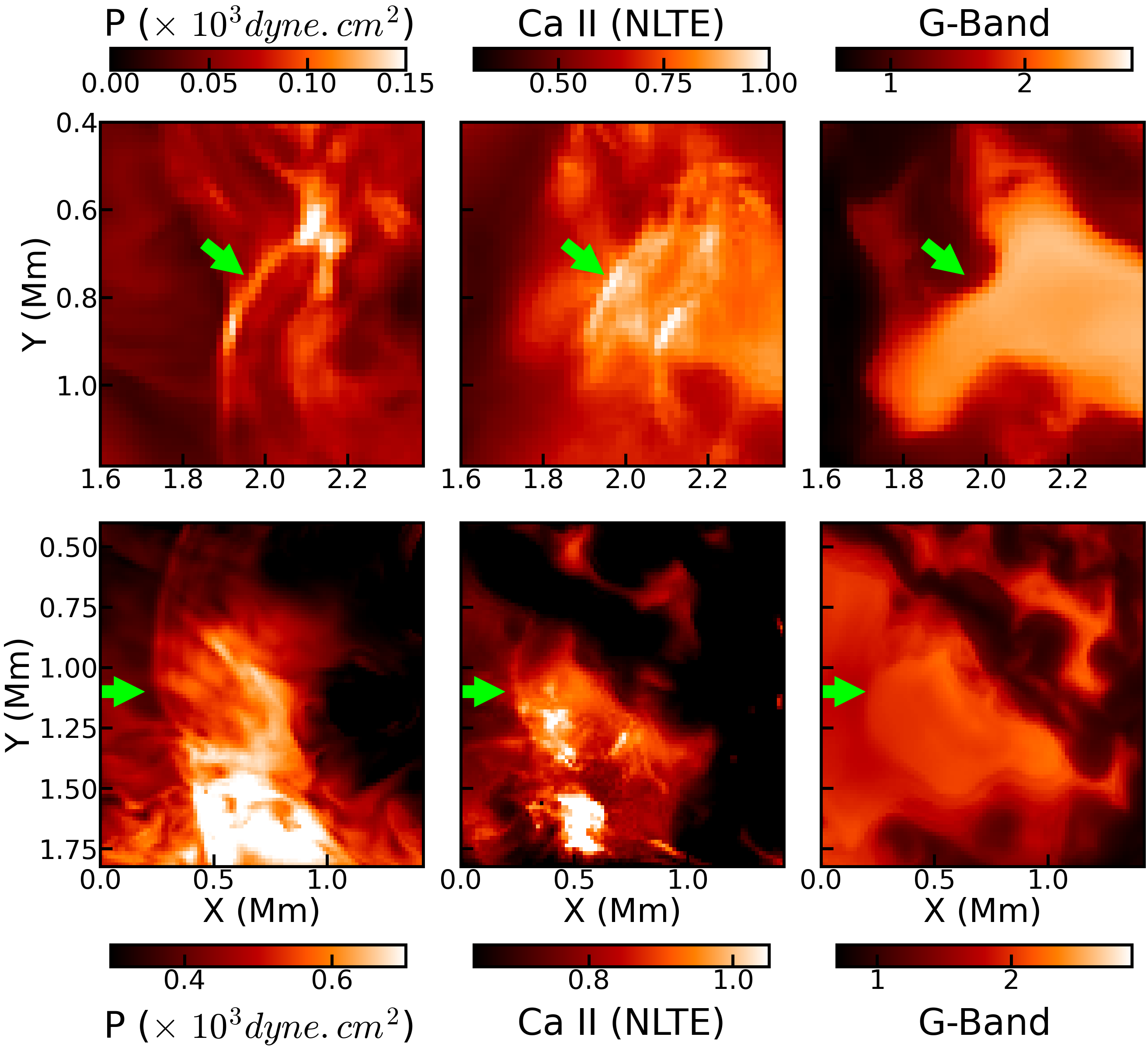}}
\caption{Two instances of a wavefront are visible in the synthesized \ion{Ca}{2} K (NLTE) channel but not in the G-band channel. Synthesized spectra for both channels are convolved with the DKIST/VBI filter profile to simulate observations. Green arrows in both rows indicate the location of the wavefronts. Due to the sources originating in the upper photosphere/chromosphere region, G-band images are delayed by 140 seconds for the chromospheric source (Top) and 58 seconds for the upper photospheric source (Bottom), accounting for the temporal delay of wave propagation in the stratified atmosphere.}  
\label{fig:figure2}
\end{figure}

\subsection{DKIST/VBI Synthesis}

\begin{figure*}[t!]
\centerline{\includegraphics[width=0.9\linewidth]{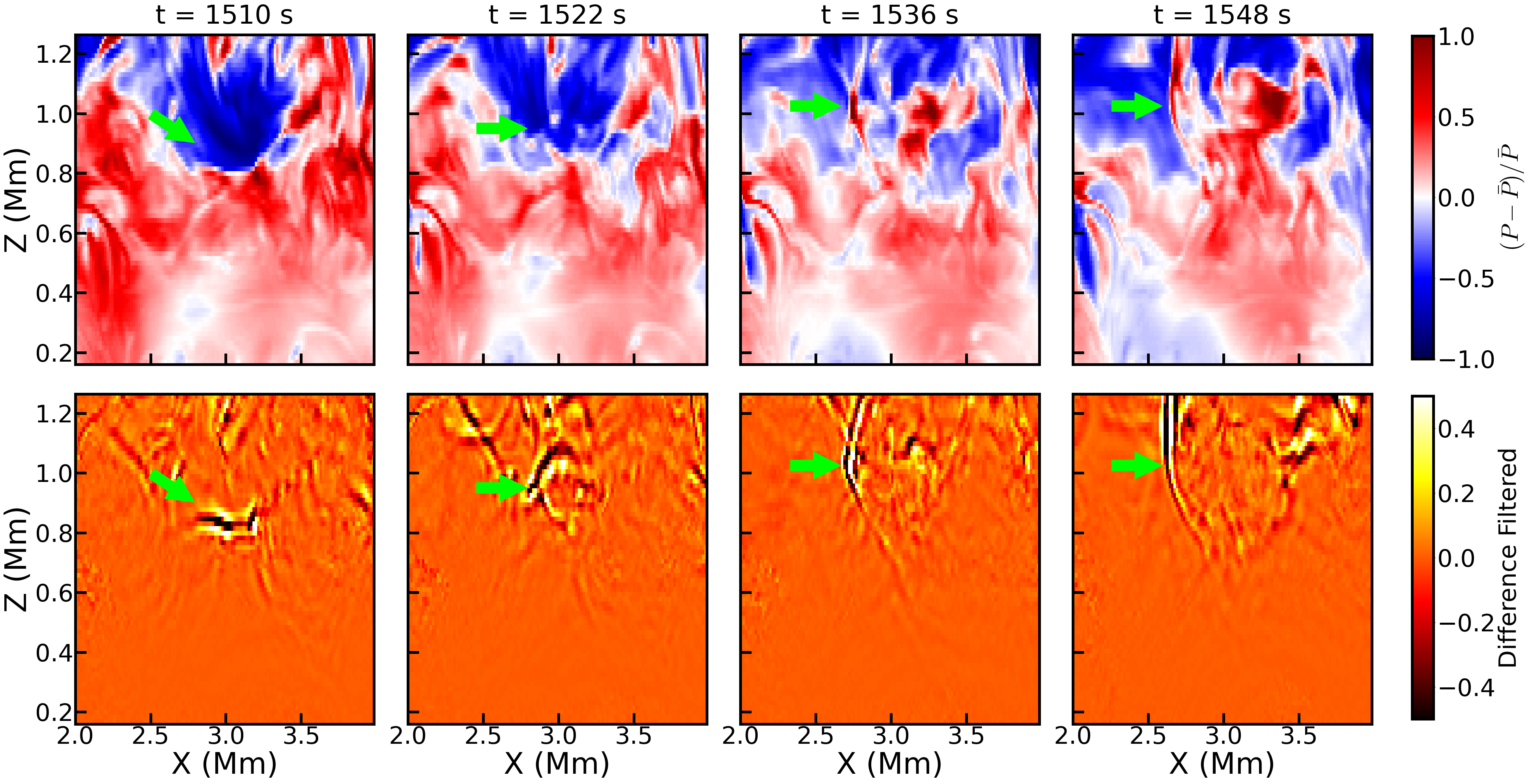}}
\caption{Local generation of a propagating wavefront approximately 900 km above the photospheric surface. The top row presents the timeseries of pressure perturbations, depicting the generation of acoustic waves by the interaction between an upward propagating shock and a chromospheric downflow. In the bottom row, a three-difference-in-time filter is applied to enhance the visibility of the locally generated wavefront~\citep{2023ApJ...955...31B}. In both rows, green arrows indicated the locations of the wavefront. Such events are quite common and appear to be an important shock dissipation mechanism in the chromosphere.}  
\label{fig:figure3}
\end{figure*}

MHD quantities such as temperature, velocity, and gas pressure, cannot be readily determined from the DKIST/VBI image timeseries. To relate the MURaM models to the observations, we constructed a forward model of the observed passbands, \ion{Ca}{2} K and G-band. Although no simultaneous continuum observations were taken, we also synthesized the DKIST/VBI blue continuum channel, as this band probes physical conditions near the base of the photosphere ($\tau_{500}=1$ in the semi-empirical models such as FALC, \cite{1993ApJ...406..319F}) and the presence or absence of wavefront signal there in the simulations may be useful in designing follow up studies.

The G-band spectra is forward modelled by the Rybicki Hummer (RH) radiative transfer code \citep{2001ApJ...557..389U}, and the blue continuum and \ion{Ca}{2} K spectra using the Spectropolarimetric NLTE inversion (SNAPI) radiative transfer code \citep{2018A&A...617A..24M}. Both codes employ the approximate lambda iteration (ALI) formalism \citep{1991A&A...245..171R} to solve for atomic level populations in non-local thermodynamic equilibrium (NLTE). The G-band wavelengths were synthesized under the LTE assumption, whereas the \ion{Ca}{2} K wavelengths were synthesized considering NLTE ionization of hydrogen and complete frequency redistribution (CRD) in the line. Both bandpasses were synthesized for a set of densely sampled wavelengths $\pm 2$ nm around the bandpass center, multiplied by the VBI filter profiles, and integrated over wavelength to emulate the observations. 

\subsection{Wavefronts and Wave Source Identification}

Analysis of the synthesized \ion{Ca}{2} K channel indicates persistent presence of many propagating circular and semi-circular wavefronts in the MURaM simulated atmosphere. As in the observations, the co-aligned synthesized G-band channel shows no sign of these near the \ion{Ca}{2} K wavefront locations or the inferred excitation sites either before or after their occurrence in the \ion{Ca}{2} K channel.  Moreover, direct investigation of the simulation output (without spectral synthesis) finds no associated photospheric wave sources and only occasional very weak associated photospheric pressure perturbations 150 s either side of the \ion{Ca}{2} K channel events. Further analysis, such as that described below for two example wavefronts, indicates that these ubiquitous wavefronts in the synthesized \ion{Ca}{2} K timeseries have, as their source, dynamical events in the lower atmosphere (at the top of the convective overshoot region) or in the simulated chromosphere proper. The wavefronts do not propagate upward from the photosphere below nor do they propagate downward from the chromosphere into the underlying photosphere to a depth that would make them visible in the G-band channel. Finally, the analysis of individual excitation sites indicates that the observed wavefronts emerge from sites of localized pressure perturbations and exhibit a wavefront speed close to the acoustic speed, although the role of magnetic field remains uncertain.

\begin{figure*}[t!]
\centerline{\includegraphics[width=0.9\linewidth]{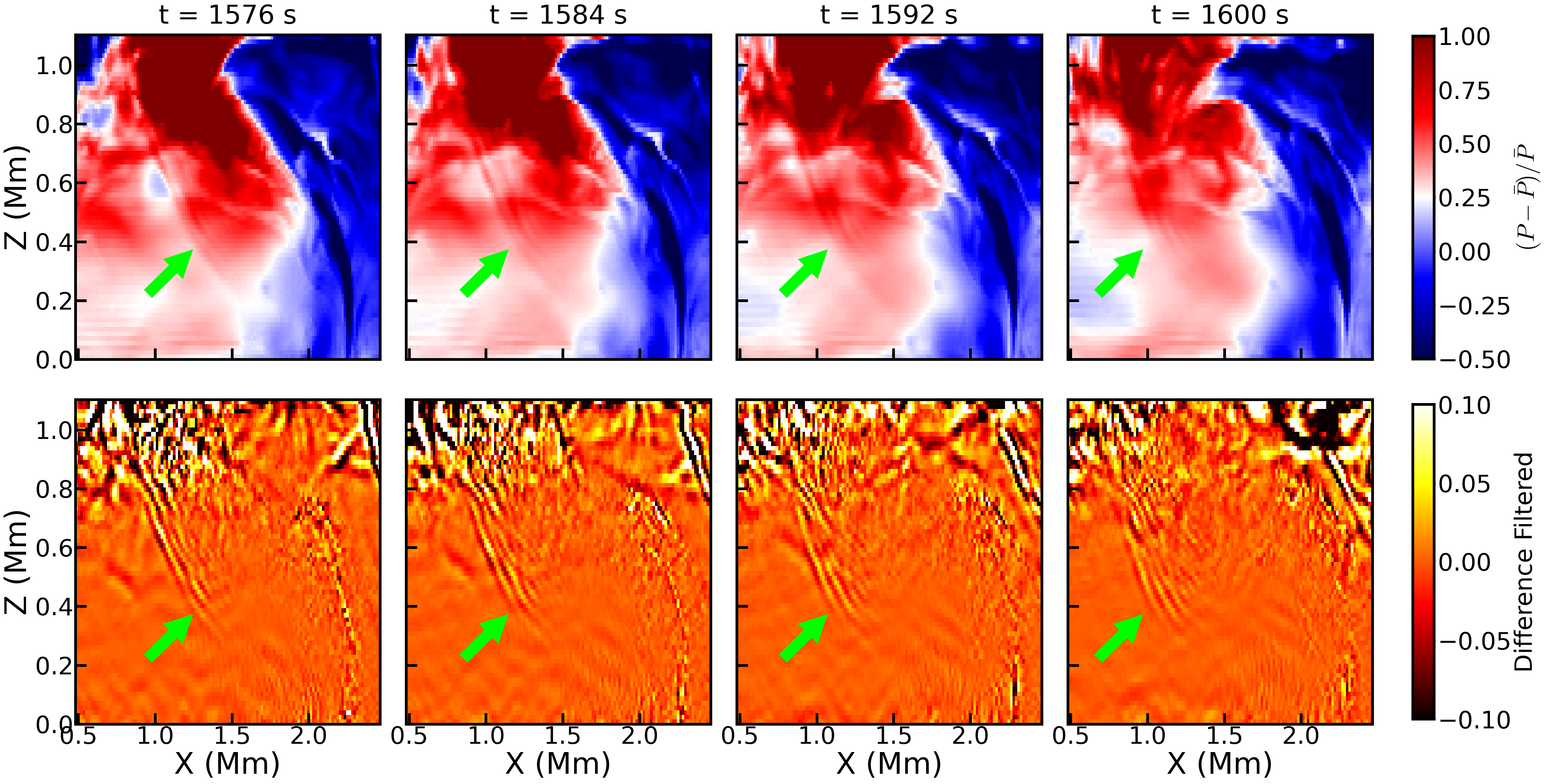}}
\caption{Generation of a acoustic wavefront approximately 400 km above the photospheric surface. The top row presents the timeseries of pressure perturbations, depicting the generation of acoustic waves from the turbulent interaction between  down-welling fluid and a large overshooting granule. In the bottom row, a three-difference-in-time filtering has been applied to enhance visibility, eliminating background contributions from the local wave perturbations~\citep{2023ApJ...955...31B}. In both rows, green arrows highlight the locations of the wavefronts.}  
\label{fig:figure4}
\end{figure*}

Figure \ref{fig:figure2}, shows snapshots of two illustrative wavefronts in the MURaM simulated atmosphere that are observationally analogous to those in the DKIST/VBI image time series. Investigation of the underlying acoustic sources suggests that these two wavefront originate with distinct physical processes. As illustrated by Figure ~\ref{fig:figure3}, the stronger and more compact wavefront (top row in Figure ~\ref{fig:figure2}) initiates with a source  at a height of about 900 to 1100 km above the photosphere. The wavefront that results is visible in both the raw pressure-fluctuation field and its temporal difference (a three-difference filter as motivated and described in~\cite{ 2023ApJ...955...31B}). The difference-filter significantly amplifies the signal and removes background contributions.

In this case, and for other sources found in the simulations at these heights, the chromospheric acoustic source event appears to be an interaction between an upward propagating shock and downward moving chromospheric plasma (See Supplementary Video 2). The initial pressure perturbation that develops into the shock can be traced back to the photosphere where it likely originates in a p-mode coherence patch. Traveling upward through the temperature minimum and into the chromosphere, the acoustic pulse experiences pronounced steepening ultimately developing into a propagating shock, until in the chromosphere, it encounters strongly downward moving plasma. The abrupt compression of the plasma in that collision gives rise to a new acoustic wavefront that becomes visible in the \ion{Ca}{2} K channel, as shown.  Importantly, as upward shock propagation is pervasive in the low MURaM atmosphere, occurring every few minutes at any given location, such events are quite common, and secondary acoustic-wave radiation as a result of this type of dynamical interaction in the chromosphere appears to be an important shock dissipation mechanism at the height being observed.

The second acoustic wavefront event illustrated in Figure \ref{fig:figure2} (bottom row) can be traced back to an origin near the temperature minimum, situated between the photosphere and the chromosphere at a height of approximately $400-600$ km. Sources at this height originate due to the dynamic interaction between upflows and downflows in the convective overshoot region, often adjacent to strong granular upflows. When the downflows encounter the upflowing granule, collision between the flows induces localized pressure fluctuations which radiate acoustic wavefronts, a Reynolds stress process reminiscent of the Lighthill mechanism~\citep{1952RSPSA.211..564L,1954RSPSA.222....1L} although the Mach number of the flow is locally only approximately 1.2. Figure \ref{fig:figure4} illustrates this, showing the pressure fluctuations in the upper row and their temporal difference in the bottom. As a strong downflow encounters a large, expanding granule, excess pressure builds up in the interface and acoustic waves are emitted in both the upward and downward directions. These wavefronts are comparatively weaker than those produced by the shock/flow interactions previously discussed. They are most easily detected in the convective overshoot region, where the background atmosphere is quite quiescent. As they propagate upward, their spatial coherence is disrupted by the more vigorous background wavefield of the low chromosphere, and as they traverse downwards, the amplitude of the waves diminishes due to the dynamic motion of granules in the photosphere (See Supplementary Video 3).



\subsection{Photospheric Sources}

Past theoretical analyses and observational studies of chromospheric waves have focused on acoustic sources in the upper convection zone of the Sun that are primarily associated with convective dynamics located in or below the solar photosphere. Many such sources can be identified in the MURaM simulations when temporal-difference filtering is employed~\citep{2021ApJ...915...36B, 2023ApJ...955...31B}.  
It is important to assess whether the waves generated by the photospheric sources propagate upward to chromospheric heights in the simulations and thus remain source candidates for at least some of the wavefronts observed in the DKIST/VBI \ion{Ca}{2} K channel.

\begin{figure}[t!]
\centerline{\includegraphics[width=\linewidth]{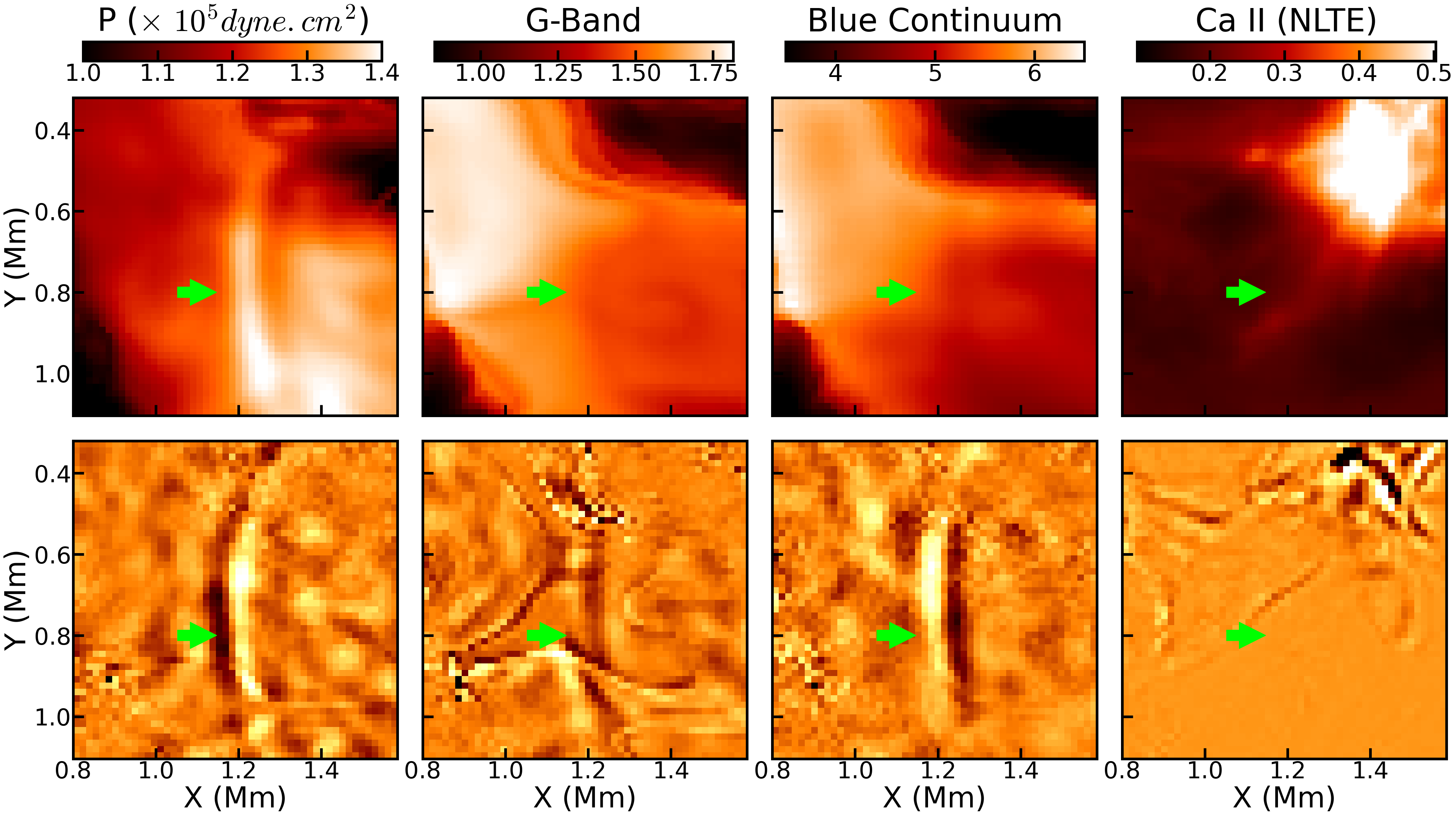}}
\caption{A local wavefront originating from a subsurface source in the simulated solar photosphere. Top Row: The wavefield is discernible in the pressure field, yet its visibility is limited in DKIST/VBI-resolution synthesized G-band, Blue-continuum, and \ion{Ca}{2} K (NLTE) channels due to weak SNR. Bottom Row: To improve visibility, three-difference temporal filtering~\citep{2021ApJ...915...36B, 2023ApJ...955...31B} is applied to de-noise the images. Even in the temporal-difference images, the wavefield remains absent in the \ion{Ca}{2} K channel. In both rows, green arrows highlight the location of the wavefronts.}  
\label{fig:figure5}
\end{figure}

Figure \ref{fig:figure5}, presents a snapshot of a wavefront produced by a sub-surface acoustic source.  Shown are the pressure perturbation in the solar photosphere and synthesized DKIST/VBI \ion{Ca}{2} K, G-band, and Blue Continuum signals along with a three-difference temporal filter of each.  Since such photospheric sources originate in the solar surface, the \ion{Ca}{2} K images shown in Figure \ref{fig:figure5} are delayed by 140 seconds, accounted for the wave travelling upward through a stratified atmosphere. Although the wavefront is visible in the MURaM pressure field, it is barely discernible in the synthesized VBI channels without filtering, and even with filtering it is not seen in the \ion{Ca}{2} K channel. Photospheric acoustic wavefronts, when of high enough amplitude, are observable in the synthesized G-band and Blue Continuum channels but not in the \ion{Ca}{2} K channel. This is precisely opposite of what the DKIST observations show, in which wavefronts are prominent in the VBI \ion{Ca}{2} K channel but not seen in G-band, and suggests that individual impulsive excitation events in the upper convection zone and solar photosphere do not produce waves that reach the heights that the DKIST VBI \ion{Ca}{2} K bandpass samples with sufficient amplitude to account for the \ion{Ca}{2} K wavefronts observed.  The difference between these individual source events and the upward propagating shocks, that do induce a secondary acoustic source in the chromosphere, lies in the fact that the latter are likely the result of p-mode coherence. Therefore, they are likely to have a significantly higher initial amplitude in the photosphere than the former.

\section{Conclusion} \label{sec:outro}

This article presented an investigation of the observational signatures and underlying sources of the lower-chromospheric transient wavefronts found in data from the DKIST/VBI observations and in MURaM simulations. Spatially compact and temporally discrete wavefronts propagating away from source sites in the solar chromosphere are visible in the \ion{Ca}{2} K DKIST/VBI channel but are absent from the co-aligned G-band channel even when travel time delays are accounted for. This suggests that their origin is in the chromosphere proper.  

We compare these observational findings with numerical simulations and confirm the persistent presence of propagating wavefronts in synthesized \ion{Ca}{2} K timeseries. Further, analysis of the simulations suggests that these waves are acoustic in nature and originate from distinct physical sources and processes.  Three such processes were identified:  upward propagation of acoustic wavefronts from sources in and below the photosphere, turbulent excitation at the top of the convective overshoot region, and upward propagating shock interaction with downward moving plasma in the chromosphere leading to acoustic radiation and shock dissipation.  Only the latter two show the characteristic \ion{Ca}{2} K presence and G-band absence of the observed transients.

The transient events we have identified have significant diagnostic potential. They promise a new helioseismic avenue for the high-resolution inference of the solar atmospheric properties at different heights~\citep{2023ApJ...955...31B}. The discovery of acoustic radiation from upward propagating shocks may be a particularly significant result of this work.  It may provide a way to directly investigate p-mode energy deposition in the lower solar atmosphere. In this regard, it is important to understand some critical differences between the observed and simulated wavefronts. Although the observed wavefronts exhibit similar SNR ($\sim 3$) against the background as those found in the simulations, they traverse a greater distance before becoming indistinguishable from it. Consequently, the observed wavefronts become longer in extent as the wavefront expands. They also show greater cross-sectional widths than those in the simulations which is related to the details of their excitation.  Together these differences suggest that wavefront sources in the Sun have larger amplitudes and durations, and/or that wave damping in the solar atmosphere is weaker than it is in the simulations.   

The observed wavefronts also propagate with a higher speed (approximately $14$ km/s) than do those in the simulation ($\sim 9$ km/s).  This may reflect difference in the magnetized nature of the real and simulated chromosphere or may be due to differences in the advection of the wave fronts by the local flows. The mean horizontal velocity amplitude in the MURaM atmosphere at the formation height of \ion{Ca}{2} K has a value of $2$ km/s, while the average speed of sound at those heights is approximately $7$ km/s. The sum of these is consistent with the observed wavefront velocity in the simulation which is about $9$ km/s. If the observed wavefront speeds are thus due to advection, there is an implied correlation between the direction of wavefront propagation and background motions. This correlation may be due to an observational bias, as the wavefronts could be more visible in areas with smoother horizontally diverging flows. Such an effect is evident in the previous analysis of photospheric wavefronts~\citep{2023ApJ...955...31B}. Interestingly, the wave propagation speeds of the slow and fast modes of magneto-acoustic waves in the MURaM low chromosphere are $6.5$ km/s and $9.2$ km/s, respectively, the latter also close to the observed wavefront velocity in the simulation. However, the role of magnetism remains uncertain in our observations since polarization measurements at this cadence are currently unavailable. Unraveling these contributions is a key goal for future research.

While conducting our study, a possible alternate structure underlying the observed wavefronts became apparent, fine-scale chromospheric fibrils ~\citep{2023ApJ...954L..35D}. Fibrils are frequently observed in the \ion{Ca}{2} K core and their motions may account for propagating narrow structures in the time series. However, the temporal dynamics of fibrils are notably slower than the features we examined, and they are not expected to exhibit expanding wavefront-like propagation from a source site at a speed close to the sound speed. It is crucial to acknowledge that amidst the background chromospheric wavefield, discerning whether a propagating front results from a localized acoustic source or arises from the spatially distributed evolving pressure field is challenging. This task is even more difficult than identifying photospheric wave-fronts, because in the chromosphere there is little time-scale separation between the background "noise" and the signal on which to base the filtering~\citep{2023ApJ...955...31B}.  

DKIST observation of wavefront propagation in the solar atmosphere promises significant advancement in our understanding of both the complex source dynamics and the background through which the source generated wavefronts travel.  With careful study, the presence of acoustic source sites in the deep solar photosphere, in the convective overshoot region, and in the lower solar chromosphere will allow ultra-local helioseismic inference of the atmospheric properties, reassessment of wave-heating and momentum transport mechanisms as a function of height, and increased confidence in our ability to model this critical region of the solar atmosphere.

\vskip0.2in
\noindent
The authors thank Han Uitenbroek for valuable guidance and advice on the synthesis of the G-band spectrum.  This work was partially supported by the National Science Foundation Award Number 2206589 and the National Solar Observatory's DKIST Ambassadors program. The National Solar Observatory is a facility of the National Science Foundation operated under Cooperative Support Agreement number AST-1400450. This research was also supported by the International Space Science Institute (ISSI) in Bern, through ISSI International Team project 502 (WaLSA: Waves in the Lower Solar Atmosphere). The simulation material is based upon work supported by the NSF National Center for Atmospheric Research, which is a major facility sponsored by the U.S. National Science Foundation under Cooperative Agreement No. 1852977. 

\vspace{5mm}
\facilities{The Visible Broadband Imager (VBI) on the National Science Foundation's Daniel K. Inoye Solar Telescope (DKIST).}



\appendix

\section{Radiative Transfer}

The MURaM atmosphere is transformed to generate a SNAPI-compatible cube for simulating the \ion{Ca}{2} K channel and an RH-compatible cube for simulating the G-band. In order to forward model \ion{Ca}{2} K channel, SNAPI resolves spectra across 601 wavelength points spanning from 393.03 nm to 393.63 nm. The synthesis of G-band channel, with its broader filter FWHM which incorporates numerous molecular and atomic lines, requires RH to forward model spectra across 3001 wavelength points ranging from 429.0 nm to 432.0 nm.  In addition, the molecular files of RH is populated appropriately to account for the absorption and scattering by the abundant molecules (such as CH) present in the solar photosphere. 

The simulation of the VBI response for each channel involves the application of a filter profile across the spectrum. The filter profiles for the \ion{Ca}{2} K and G-band channels are modeled using the equation: profile $= \frac{1}{1 + t}$ where $t = \left( \left| \lambda - \lambda_0 \right| / \frac{\Delta}{2} \right)^{2L}$. Here, $\lambda_0$ is the center wavelength, $\Delta$ is the Full Width Half Maximum (FWHM) of the channels, and $L$ = cavity length ($L = 2$ is used for both \ion{Ca}{2} K and for G-band).




\bibliography{acoustic}{}
\bibliographystyle{aasjournal}



\end{document}